\documentclass[
    ,final            % use final for the camera ready runs
  ]
  {aipproc}

\layoutstyle{8x11single}

\usepackage{amsmath,latexsym,amssymb}
\usepackage{theorem}

\newcommand{\nc}{\newcommand}
\nc{\rnc}{\renewcommand} \nc{\beq}{\begin{equation}}
\nc{\eeq}{{\end{equation}}} \nc{\bea}{\begin{eqnarray}}
\nc{\eea}{\end{eqnarray}} \nc{\beqa}{\begin{eqnarray}}
\nc{\eeqa}{\end{eqnarray}} \nc{\lbar}[1]{\overline{#1}}
\nc{\bra}[1]{\langle#1|} \nc{\ket}[1]{|#1\rangle}
\nc{\ketbra}[2]{|#1\rangle\!\langle#2|}
\nc{\braket}[2]{\langle#1|#2\rangle} \nc{\proj}[1]{|
#1\rangle\!\langle #1 |} \nc{\avg}[1]{\langle#1\rangle}
\rnc{\max}{\operatorname{max}} \nc{\Rank}{\operatorname{Rank}}
\nc{\smfrac}[2]{\mbox{$\frac{#1}{#2}$}}
\nc{\Tr}{\operatorname{Tr}} \nc{\ox}{\otimes} \nc{\dg}{\dagger}
\nc{\dn}{\downarrow} \nc{\cA}{{\cal A}} \nc{\cB}{{\cal B}}
\nc{\cC}{{\cal C}} \nc{\cD}{{\cal D}} \nc{\cE}{{\cal E}}
\nc{\cF}{{\cal F}} \nc{\cG}{{\cal G}} \nc{\cH}{{\cal H}}
\nc{\cI}{{\cal I}} \nc{\cJ}{{\cal J}} \nc{\cK}{{\cal K}}
\nc{\cL}{{\cal L}} \nc{\cM}{{\cal M}} \nc{\cN}{{\cal N}}
\nc{\cO}{{\cal O}} \nc{\cP}{{\cal P}} \nc{\cR}{{\cal R}}
\nc{\cS}{{\cal S}} \nc{\cT}{{\cal T}} \nc{\rU}{{\cal U}}
\nc{\cX}{{\cal X}} \nc{\cZ}{{\cal Z}}
\nc{\csupp}{{\operatorname{csupp}}}
\nc{\qsupp}{{\operatorname{qsupp}}} \nc{\var}{\operatorname{var}}
\nc{\rar}{\rightarrow} \nc{\lrar}{\longrightarrow}
\nc{\poly}{\operatorname{poly}}
\nc{\polylog}{\operatorname{polylog}} \nc{\1}{\openone}

\def\>{\rangle}
\def\<{\langle}

\def\a{\alpha}
\def\b{\beta}

\def\e{\epsilon}

\def\h{\eta}

\def\l{\lambda}

\def\r{\rho}

\nc{\glneq}{{\raisebox{0.6ex}{$>$}  \hspace*{-1.8ex}
\raisebox{-0.6ex}{$<$}}}
\nc{\gleq}{{\raisebox{0.6ex}{$\geq$}\hspace*{-1.8ex}
\raisebox{-0.6ex}{$\leq$}}}

\nc{\RR}{{{\mathbb R}}} \nc{\CC}{{{\mathbb C}}} \nc{\FF}{{{\mathbb
F}}} \nc{\NN}{{{\mathbb N}}} \nc{\ZZ}{{{\mathbb Z}}}
\nc{\PP}{{{\mathbb P}}} \nc{\QQ}{{{\mathbb Q}}} \nc{\UU}{{{\mathbb
U}}} \nc{\WW}{{{\mathbb W}}} \nc{\EE}{{{\mathbb E}}}
\rnc{\SS}{{{\mathbb S}}} \nc{\id}{{\operatorname{id}}}

\begin{document}

\title{Entanglement in Random Subspaces}

\author{Patrick Hayden}{
  address={School of Computer Science \\
  McGill University \\
  3480 University St., McConnell Engineering Building, Rm. 318 \\
  Montreal, Quebec, Canada H3A 2A7}
}

%\author{<author2>}{
%  address={<common address for author2 and author3>}
%}

%\author{<author3>}{
%  address={<common address for author2 and author3>}
%  ,altaddress={<author1 address>} % additional visiting address
%}

\begin{abstract}
The selection of random subspaces plays a role in quantum
information theory analogous to the role of random strings in
classical information theory. Recent applications have included
protocols achieving the quantum channel capacity and methods for
extending superdense coding from bits to qubits. In addition,
random subspaces have proved useful for studying the structure of
bipartite and multipartite entanglement.
\end{abstract}

\maketitle

%%%%%%%%%%%%%%%%%%%%%%%%%%%%%%%%%%%%%%%%%%%%
%% MAINMATTER
%%%%%%%%%%%%%%%%%%%%%%%%%%%%%%%%%%%%%%%%%%%%

\section{Introduction}

In quantum information theory, we're fond of saying that Hilbert
space is a big place, the implication being that there's room for
the unexpected to occur. You will receive no counterarguments to
that homespun wisdom from me for the simple reason that, as far as
I can tell, it's correct. I'm going to present a number of results
in quantum information theory that stem from the initially
counterintuitive geometry of high-dimensional vector spaces, where
subspaces with highly extremal properties are the norm rather than
the exception. Peter Shor has shown, for example, that randomly
selected subspaces can be used to send quantum information through
a noisy quantum channel at the highest possible rate, that is, the
quantum channel capacity~\cite{S02}. More recently, Debbie Leung,
Andreas Winter and I demonstrated that a randomly chosen subspace
of a bipartite quantum system will likely contain \emph{nothing}
but nearly maximally entangled states, even if the subspace is
nearly as large as the original system in qubit
terms~\cite{HLW04}. This observation has implications for
communication, especially superdense coding, as described
in~\cite{AHSW04}. Those results and the intuition behind them will
be my focus here.

\section{Surprises in high dimension}

Suppose, for the moment, that you are an astronaut orbiting Earth
in a space shuttle. Imagine, slightly less plausibly, that you are
a also mathonaut, meaning that you observe not our Earth but
instead a highly idealized version of it in which the population
is evenly distributed over the whole surface of the planet. Bored
with the daily routine of gyroscope failures and rebreather
malfunctions, you decide to look out the window and count the
number of people living within a ten kilometer band of the
equator. (You have both a good telescope and lots of time on your
hands.) Give or take a few, you find five million people, with the
rest of the population of six billion living elsewhere. Now, bold
mathonaut that you are, you repeat your observations in higher and
higher dimensions, first counting the inhabitants of a ten
kilometer thickening about the equator of a 3-sphere version of
the Earth (in four dimensions), then of a 4-sphere and so on up.
Long before you reach the 50-sphere, however, you discover a great
time saver: count the people \emph{outside} of the band. There
aren't any. Perplexed, you decide to check if your luck was bad by
selecting other equators for the 50-sphere, but each time you find
that every single inhabitant of the planet lives within ten
kilometers of the equator.

What's going on? Nothing too sophisticated, it turns out. The
calculation itself is completely elementary, an exercise in
spherical coordinates, but the effect is an example of the broader
``concentration of measure'' phenomenon: naturally defined random
variables on high-dimensional spaces tend to concentrate strongly
around their average values~\cite{L01}. The most familiar example
of this is probably the case of the sum of $n$ independent,
bounded random variables. According to Chernoff's bound, the
probability that the sum deviates more than $\epsilon$ from its
mean value is less than $\exp(-Cn\epsilon^2)$ for some positive
constant $C$. The analogous statement for functions on the
$k$-sphere is known as Levy's Lemma:

\medskip

\noindent {\bf Levy's Lemma} \emph{(See \cite{MS86}, Appendix IV,
and \cite{L01}) \label{lem:Levy} Let $f : \SS^{k} \rar \RR$ be a
function with Lipschitz constant $\h$ (with respect to the
Euclidean norm) and a point $X \in \SS^{k}$ be chosen uniformly at
random. Then}
\begin{equation}
\Pr\left\{ f(X) - \bar{f} ~ \glneq ~ \pm\a \right\}
  \leq \exp\left( -C (k-1) \a^2 / \h^2 \right)
\end{equation}
\emph{for some constant $C>0$.}

\medskip

\noindent Here $\bar{f}$ is used to denote either the mean value
or a median for $f$; the median is actually a more natural
quantity in the theory of concentration of measure. The function
relevant to our mathonaut investigations is simply
$f(x_1,\ldots,x_n)
= x_1$, which obviously has Lipschitz constant one and both mean
and median of zero.

\section{Random states and random subspaces}

Quantum states, of course, are represented as unit vectors, so
Levy's Lemma provides a ready-made tool for exploring the
properties of random quantum states in high-dimensional systems.
We need only choose the function $f$ and plug in its mean value.

The example that will occupy us is the entanglement of a bipartite
system. Let $\ket{\phi}$ be a random pure state in $\CC^{d_A} \ox
\CC^{d_B}$, chosen according to the unitarily invariant measure,
which in turn corresponds to the uniform measure on the $(2 d_A
d_B - 1)$-sphere. Assuming without loss of generality that $d_A
\leq d_B$, the expected value of the entropy of entanglement is
known~\cite{LP88,P93,FK94,S-R95,S96} and satisfies
\begin{equation}
\EE E(\phi) = \EE S(\phi_A) \geq \log_2 d_A - \frac{d_A}{2 \ln 2
d_B},
\end{equation}
where $S$ is the von Neumann entropy. Since any state of this
bipartite system can have no more than $\log_2 d_A$ ebits of
entanglement, this tells us that on average the entanglement is
within one ebit of being maximal. Levy's Lemma allows us to
quantify how likely it is that the entanglement of a random state
will fall significantly below the mean. Define $\b
= \smfrac{1}{\ln 2}\smfrac{d_A}{d_B}$. Once all the calculations
are done, we get the following bound:
\begin{equation} \label{eqn:singleState}
 \Pr\left\{ S(\phi_A) < \log_2 d_A - \a - \b \right\}
 \leq
 \exp\left( - \frac{(d_A d_B - 1)C \a^2}{(\log d_A)^2} \right),
\end{equation}
for some $C > 0$ provided $d_B \geq d_A \geq 3$. Ignoring the
small $(\log d_A)^2$ factor in the denominator of the exponent,
this is the same type of exponential convergence to the mean that
occurs for population evenly distributed on the $k$-sphere.

The convergence is so rapid, in fact, that it is possible to
strengthen these results about random states into statements about
random subspaces. The idea is to fix a subspace $S_0$ of dimension
$s$ and choose a very fine net of states $\cN_0 \subset S_0$, so
fine that given any state $\ket{\phi} \in S_0$, there is an
approximating $\ket{\tilde{\phi}} \in \cN_0$ such that $\| \phi -
\tilde{\phi} \|_1 \leq \e$. If we choose a random unitary $U$
according to the Haar measure, it takes $S_0$ to a random subspace
$U S_0$ and it takes the net $\cN_0$ to a net $U \cN_0$ for the
new subspace. The probability that a given state in $U \cN_0$ has
entanglement less than $\log_2 d_A - \a - \b$ is given by
Equation~(\ref{eqn:singleState}) while the probability that any
one of them has entanglement less than $\log_2 d_A - \a - \b$ is
then bounded above by
\begin{equation} \label{eqn:manyStates}
 |\cN_0| \exp\left( - \frac{(d_A d_B - 1)C \a^2}{(\log d_A)^2}
 \right).
\end{equation}
As a net on the unit ball of a subspace of real dimension $2s$,
the size of $\cN_0$ will scale as $(C/\e)^{2s}$ for some constant
$C > 0$. Proving the existence of a subspace in which all states
are highly entangled then becomes a matter of tuning the
resolution of the net $\cN_0$ and the value of $\a$. We find that
when $d_B \geq d_A \geq 3$ and $0
< \a
< 1$, there exists a subspace of $\CC^{d_A} \ox \CC^{d_B}$
of dimension
\begin{equation}
\left\lfloor d_A d_B \frac{C \a^{2.5}}{(\log d_A)^3}
\right\rfloor,
\end{equation}
where $C >0$ is, as always, a constant. From now on, I'll refer to
a subspace having this property (for fixed $\a$) as a
\emph{maximally entangled subspace}. In qubit terms, in a
bipartite system of $n$ by $n + o(n)$ qubits, this is a subspace
of size $2n - o(n)$ qubits in which \emph{all} of the states of
entanglement at least $n - o(1)$ ebits. The maximally entangled
subspace is nearly as large as the whole system.

For the sake of unfair comparison, we could consider the subspace
spanned by any two Bell states of a pair of qubits. Any such
subspace will not only fail to contain only nearly maximally
entangled states, it will always contain some product states!

\section{Consequences for entanglement measures}

Should we be alarmed? Unconcerned? Unconvinced? One way to place
the result in context is to reinterpret it in the language of
mixed state entanglement measures. Consider the maximally mixed
state $\r$ on one of the maximally entangled subspaces. Because
the range of $\r$ consists only of these states, any convex
decomposition of $\r$ into pure states will again be into these
nearly maximally entangled states. Continuing to use the language
of qubits, in an $n$ by $n+o(n)$ qubit system, $\r$ will have
entanglement of formation
\begin{equation}
E_f(\r) = n - o(1),
\end{equation}
which is nearly maximal. On the other hand, as the maximally mixed
state on a subspace of $2n - o(n)$ qubits, $\r$ will have entropy
at least $S(\r) = 2n - o(n)$. In fact, the parameters can be tuned
such that the quantum mutual information satisfies
\begin{equation}
S(\r_A) + S(\r_B) - S(\r_{AB}) = O(\log n).
\end{equation}
The distillable entanglement of the $\r$ is therefore also $O(\log
n)$~\cite{VP98}. This leaves a huge gap between the entanglement
of formation and the entanglement of distillation, the first being
almost as large as it can be with the second simultaneously nearly
as small as it can be. Ignoring issues of the additivity of the
entanglement of formation, the state $\r$ provides an example of a
state that is nearly as hard to make as a maximally entangled
state and yet is nearly useless as a resource. In other words,
this $\r$ would be an example of a state exhibiting near-maximal
irreversibility.

\section{Superdense coding of quantum states}

Another way to place the existence of these maximally entangled
subspaces in context is to study their applications to
communication. Suppose that Bob has in mind a specific state
$\ket{\phi}$ on a quantum system $S$ that he would like to send to
Alice. (Their roles are reversed from the usual convention in
order to be consistent with the rest of the paper.) If $S$ were a
bipartite system $\CC^{d_A} \ox \CC^{d_B}$ and $\ket{\phi}$ was
promised to be maximally entangled, then Bob could take advantage
of superdense coding~\cite{BW92}: Alice and Bob would pre-share a
fixed maximally entangled state and in order to send $\ket{\phi}$,
Bob would apply a local unitary transformation $U_{\phi}$ before
sending his half of the system to Alice. That's fine, of course,
but the promise that $\ket{\phi}$ be maximally entangled would
seem to make this a very special case, especially since Alice ends
up with both halves of the bipartite system. Actually, thanks to
the existence of a maximally entangled subspace, this is
essentially the \emph{general} case. If Alice and Bob pre-share a
fixed maximally entangled state and agree on an embedding $S
\subset \CC^{d_A} \ox \CC^{d_B}$ of a maximally entangled
subspace, then Bob can send to Alice any state $\ket{\phi} \in S$
using the simple protocol, up to small errors, since they are all
nearly maximally entangled. The qubit accounting then works as
follows: Bob can send Alice an arbitrary $2n - o(n)$ qubit state
by consuming $n$ ebits of entanglement and sending $n+o(n)$
qubits, achieving the two-for-one savings normally associated with
sending only classical information. (An earlier version of the
result achieved the same rates but consumed extra shared random
bits~\cite{HHL03}.)

The superdense coding idea can be pushed even further, to the case
where the state to be prepared in Alice's lab is entangled with
Bob's system and, therefore, no longer pure. A quick check of the
extremal situation suggests that this should be easier: if the
goal is prepare a fixed maximally entangled state between Alice
and Bob's labs, then, provided Alice's system is no larger than
Bob's, no communication is required at all; Bob need just perform
an appropriate local unitary on his own system. The interpolation
between the two-for-one of pure states and the no communication of
maximally entangled states is analyzed in~\cite{AHSW04} using
techniques similar to those discussed here, with the result that
the largest Schmidt coefficient $\l_{\max}$ of all the states to
be prepared controls the trade-off. To leading order in the
asymptotics, $\smfrac{1}{2} \log_2 s + \smfrac{1}{2} \log_2
\l_{\max}$ qubits and $\smfrac{1}{2} \log_2 s - \smfrac{1}{2}
\log_2 \l_{\max}$ ebits are required. $s$ is defined as before to
be the dimension of the system being prepared on Alice's side
alone.

\section{Multipartite entanglement}

The results on bipartite entanglement extend easily to the
multipartite realm. For convenience, consider a random state of
$n$ qudits, so that $\ket{\phi} \in (\CC^d)^{\ox n}$ and assume
that $n$ is held fixed while $d$ is allowed to increase. The
following conclusions about random states are essentially
corollaries of what we've already seen:
\begin{itemize}
\item
 The pure state entanglement across \emph{every} bipartite cut is
 likely to be near maximal simultaneously.
\item
 If $k > n/2$ then the reduced state of any $k$ qudits will likely
 have near-maximal entanglement of formation. Meanwhile, if $k <
 n/2$ then it is likely that the entanglement of formation becomes
 less than any positive constant.
\item
 With the participation of the remaining $n-2$ parties, any pair
 of parties can distill a nearly maximally entangled pure state.
\end{itemize}
The last item is at first glance probably the most surprising but
no harder to prove than the others. The distillation protocol
consists of the remaining $n-2$ parties each measuring in a random
local basis. The state shared by the other two conditioned on the
outcome of this measurement is essentially random and, therefore,
nearly maximally entangled.

\section{Conclusion}

While it is in retrospect no surprise that techniques for dealing
with random subspaces should prove useful in quantum information
theory, the ease with which they can be analyzed was certainly a
surprise to \emph{me}. Random subspace techniques have been a
mainstay of the ``local theory of Banach spaces'' ever since
Milman~\cite{M71} gave a proof of Dvoretsky's Theorem~\cite{D61}
using concentration of measure ideas. It is amusing and perhaps
instructive to note that the title of a classic book by Milman and
Schechtman on the subject, ``Asymptotic theory of finite
dimensional normed spaces,'' concisely sums up in mathematical
terms one of the main goals of quantum information theory.

%%%%%%%%%%%%%%%%%%%%%%%%%%%%%%%%%%%%%%%%%%%%%%%%
%% BACKMATTER
%%%%%%%%%%%%%%%%%%%%%%%%%%%%%%%%%%%%%%%%%%%%%%%%

\begin{theacknowledgments}
The work presented here comes from papers I have had the pleasure
of working on with my co-authors Anura Abeyesinghe, Debbie Leung,
Graeme Smith and Andreas Winter. Also, I would like to thank the
Sherman Fairchild Foundation and the NSF, through grant number
EIA-0086038, for their support of this research.
\end{theacknowledgments}

%% If the bibliography is produced without BibTeX comment out the

\bibliographystyle{aipprocl} % if natbib is missing

\bibliography{talk}

\end{document}